\documentclass[final,authoryear,5p,times,twocolumn]{elsarticle}

\usepackage{graphics}

\usepackage{amssymb}
\usepackage{epigraph}

\usepackage{lineno}

\journal{Planetary and Space Science}

\begin{document}

\twocolumn[
\section*{\Large {\mdseries Editorial}}
\section*{\LARGE {\sc Cosmic Dust \MakeUppercase{\romannumeral 7}}}
\begin{flushleft}
\hrulefill
\end{flushleft}
\vspace{0.3in}
]

{\sl \epigraph{For dust you are and to dust you will return.}{Genesis 3:19}}

\section{Interdisciplinary research on cosmic dust}


Cosmic dust has long been recognized to play a key role in astronomy, especially in planetary science, space science, astrophysics, astrochemistry, astrobiology, and astromineralogy. 
The reason for its crucial importance is that it has been found essentially in all the environments where the temperature is low enough for gases to condense into solids. 
Dust carries critical information on the medium where it is observed and acts as a fantastic probe into the physical properties and chemical conditions of various astronomical objects.
Over the years, each astronomical community has naturally developed its own observing strategies, laboratory equipment, models, numerical codes, etc. 
As a consequence, it has become a real challenge of the present day to get a consistent and global picture of such considerably diverse approaches using cosmic dust as a tool, despite the broadly shared interest in it.

A series of Cosmic Dust meetings\footnote{{URL:} {https://www.cps-jp.org/\~{}dust}} provides a unique opportunity to overcome this difficulty. 
These meetings cover the widest range of cosmic dust research in the world by gathering experts from different fields in astronomy with the goal to share their best knowledge of dust. 
Since 2006, we have been organizing the Cosmic Dust meetings in Asia, every time with a small number of invited experts from various fields of dust, together with a limited number of contributed experts. 
The basic format of the meeting is made up of 50--60 selected attendees and ample time for both scientific discussion and the development of interpersonal relationships during long coffee breaks and enjoyable evening events. 
A banquet and a half-day excursion also contribute to making these meetings a pleasant moment, and providing a very specific place to trigger new projects and come up with emerging new ideas. 
Therefore, the overall atmosphere of the meetings is conducive to establishing long-term relationships and possible collaborations across scientific disciplines.

\begin{figure*}[t]
 \begin{center}
  \includegraphics[width=180mm]{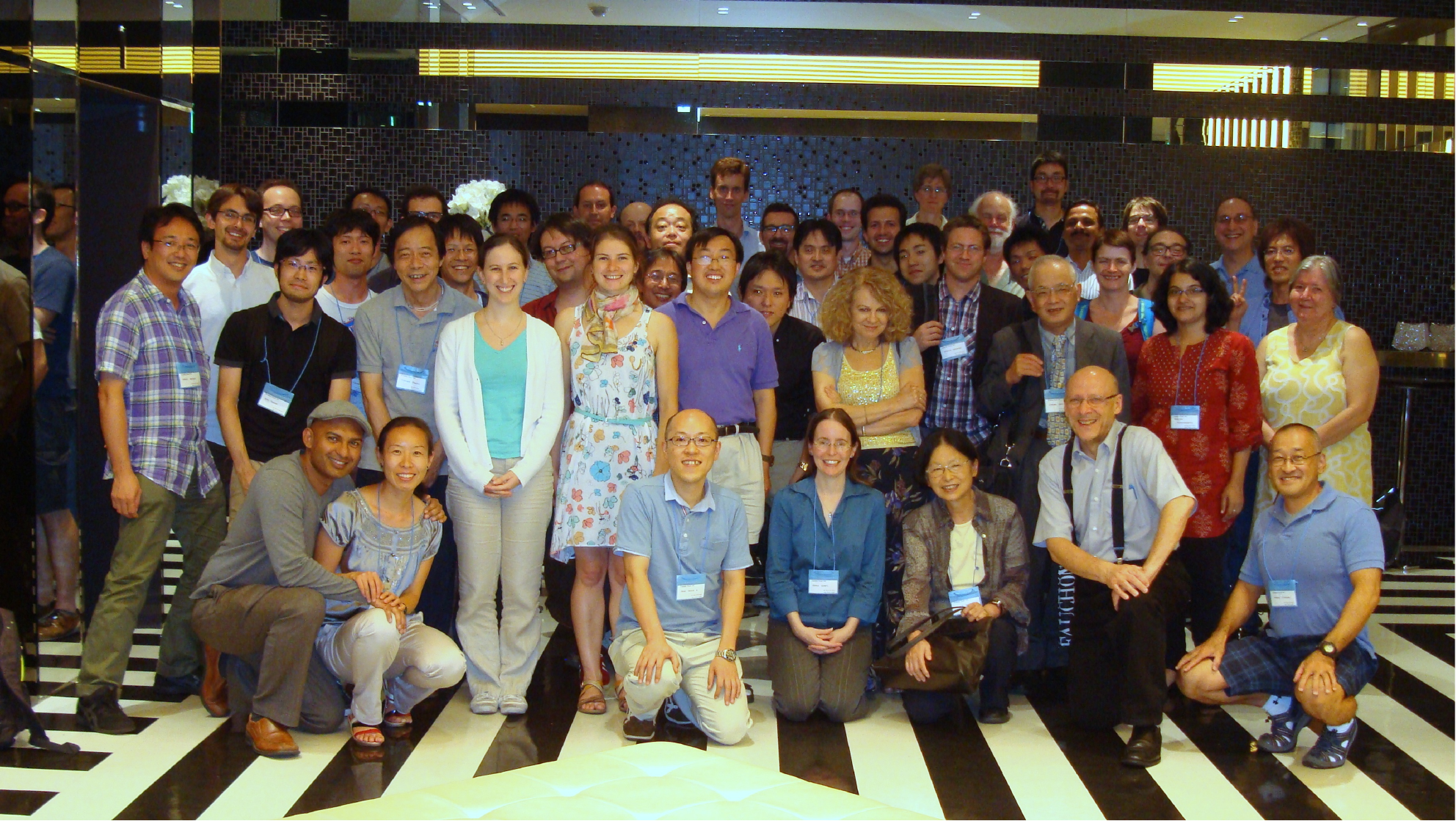}
 \end{center}
 \caption{A group picture of participants to {\sc Cosmic Dust \MakeUppercase{\romannumeral 7}}; (In no particular order) Z. Wahhaj, B. Yang, A.K. Inoue, E. Gibb, C. Koike, A.M. Magalh\~{a}es, H. Chihara, H. Senshu, R. Tazaki, T. Onaka, A. R\'{e}my-Ruyer, M. Khramtsova, A. Li, L. Kolokolova, C. Kaito, P. Shalima, J.-F. Gonzalez, G. Chiaki, S. Mori, M. Hammonds, T. Yada, T. Shimonishi, S. Bromley, R. Mason, H. Kimura, T. Birnstiel, T. Nozawa, T. Hendrix, M. Yamagishi, F. Galliano, M. Juvela, K. Murakawa, J. Olofsson, E. Bron, K. Bekki, M.S.P. Kelley, G. Aniano, T. Kokusho, V. Sterken, M.F. A'Hearn, Y. Hattori, H. Kobayashi, H.S. Das, A.C. Bell, and A. Brieva.}
 \label{fig:one}
\end{figure*}
The 7th meeting on Cosmic Dust (hereafter ``{\sc Cosmic Dust \MakeUppercase{\romannumeral 7}}'') was held at the Umeda Satellite Campus of Osaka Sangyo University, Japan, from August 4 to 8, 2014. 
It was fortunate for us that we could have a great list of invited speakers year after year from a broad range of research fields.
The invited speakers of {\sc Cosmic Dust \MakeUppercase{\romannumeral 7}} were Michael F. A'Hearn (University of Maryland, USA), Til D. Birnstiel (CfA, USA), Erika L. Gibb (University of Missouri-St. Louis, USA), Mika J. Juvela (University of Helsinki, Finland), Michael S. P. Kelley (University of Maryland, USA), Antonio Mario Magalh\~{a}es (University of S\~ao Paulo, Brazil), Rachel Mason (Gemini Obs., USA), Takaya Nozawa (NAOJ, Japan), Johan Olofsson (MPIA, Germany), Takashi Onaka (University of Tokyo, Japan), Aur\'{e}lie R\'{e}my-Ruyer (CEA Saclay, France), and Toru Yada (ISAS/JAXA, Japan).
They gave outstanding 40-min~talks covering various themes as described in the next section, and served as excellent chairpersons and members of the panel for the best poster contest.
The Best Poster Award of {\sc Cosmic Dust \MakeUppercase{\romannumeral 7}} was received by Shoji Mori for his paper entitled ``Interplay between dust and MHD turbulence in protoplanetary disks: Electric-field heating of plasmas and its effect on the ionization balance of dusty disks'' \citep{mori-okuzumi2014}.
Successful contributed speakers presented their recent work through either a 20-min~talk or a poster with a 1-min~introductory talk prior to the first poster session.
The posters were displayed throughout the meeting for intensive discussion along with extended coffee breaks that lasted for 1~h every time following each 1-h oral session.
The abstracts of the talks and the posters presented at the meeting are available for download at the Cosmic Dust website ({https://www.cps-jp.org/\~{}dust/Program\_VII.html}). 
Fig.~\ref{fig:one} is a group shot of the participants taken immediately after the banquet, held on the Dojima waterfront of the Aqua Metropolis Osaka.

This special issue in Planetary and Space Science (PSS) is aimed at collecting the new results and original ideas that were presented at {\sc Cosmic Dust \MakeUppercase{\romannumeral 7}} in 2014. 
We are pleased that this issue also includes relevant papers from scientists outside the meeting attendees, so that it warrants total coverage of the most recent findings \citep{krelowski2015,moores-et-al2015}.
In Section~\ref{review}, we shall overview the papers presented at {\sc Cosmic Dust \MakeUppercase{\romannumeral 7}}, since the reader is not necessarily familiar with the subject of this volume.
The papers are classified into subsections according to their dusty environments in the same way as the program of the meeting, although we admit that the classification is not unique.
Section~\ref{perspectives} briefly describes our perspectives for the development of cosmic dust research and ends with our concluding remarks.

\section{The contents of C{\sc osmic} D{\sc ust} \MakeUppercase{\romannumeral 7}}
\label{review}

\subsection{Solar system}

Comets are the oldest fossils of primitive ices and dust at the early stages of the Solar System formation, while their surfaces are processed by solar radiation \citep{whipple1950}.
In a classical picture of comets, the sublimation of water ice near the surface of a nucleus in the inner Solar System is a driving force for water molecules to drag dust from the surface and accelerate it until it decouples in the outer coma.
Michael F. A'Hearn reported that the picture has drastically changed within the last ten years, owing to the advent of infrared space telescopes, space missions to comets, and advances in the theory of Solar System formation \citep{ahearn2014}.
It turns out that water molecules are released primarily in the coma from large chunks of nearly pure ice and do not drag dust from the surface \citep{ahearn-et-al2011}.
In addition, the chunks, which are most likely fluffy porous aggregates of $1~\mu$m water ice grains, are found to exist separately from dark material \citep{protopapa-et-al2014}.
This is totally dissimilar from a model of icy grains proposed by \citet{greenberg1998}, in which water ice covers dark refractory grains.
The separation of icy grains and non-icy, refractory grains has also been observed in a protoplanetary disk as a temporal variation in the 3~$\mu$m water ice absorption against a constant dust continuum \citep{terada-et-al2007}.
Therefore, recent observations suggest that icy grains and refractory grains are separately present in protoplanetary and debris disks, as well as comets.
{\sl It is clear that these findings shed new light on the formation of icy grains, if the majority of water ice in comets does not form on the surfaces of refractory grains.}

Other papers covered the following topics: infrared spectral fittings of dust in cometary comae by Michael S.P. Kelley; the curation and characteristics of Hayabusa-returned particles from the S-type Near-Earth Asteroid 25143 Itokawa by Toru Yada; sub-millimeter observations of comet dust by Bin Yang; numerical simulation on the motion of electrically charged grains above the surfaces of airless bodies by Hiroki Senshu \citep{senshu-et-al2015}; a laboratory experiment on the infrared spectra of LIME (Low Iron Manganese Enriched) olivine by Hiroki Chihara; radiative transfer modeling of impact ejecta curtains by Shalima Puthiyaveettil \citep{puthiyaveettil-et-al2015}; numerical simulation on light scattering by porous spheroidal particles with rough surfaces by Himadri Sekhar Das \citep{kolokolova-et-al2015}.

\subsection{Debris disks}

Infrared space observatories such as Spitzer and Herschel have provided valuable spectral data of debris disks at different stages of evolution.
Johan Olofsson addressed dust mineralogy in debris disks as well as protoplanetary disks in terms of disk evolution and planet formation \citep{olofsson-et-al2014}. 
The spectral decomposition of the data relies on, to a great extent, laboratory measurements of infrared spectra with synthesized or natural terrestrial grains.
Among the emission features of crystalline olivine, the 69~$\mu$m feature has been used as an indicator of the grain temperature and the iron fraction in olivine grains \citep[e.g.,][]{koike-et-al2006}.
It is, however, worth mentioning that the spectral decomposition of debris disks often does not incorporate the knowledge of dust mineralogy acquired through in situ and laboratory analyses of real dust samples from comets, despite the similarity of their infrared spectra.
Laboratory measurements of the 69~$\mu$m feature were performed with pure olivine grains, but the absence of pure olivine grains is a consensus on dust mineralogy in comets.
In-situ measurements of element abundances for dust in Comet 1P/Halley revealed that olivine is always associated with organic refractory material without exception \citep{jessberger-et-al1988}.
Laboratory analyses of dust particles captured from Comet 81P/Wild 2 have revealed the presence of minerals as well as high-mass polycyclic aromatic hydrocarbon (PAH) associated with whole particles, despite the fact that organics suffered from destruction and loss during the capture and the analyses \citep{brownlee-et-al2006,sandford-et-al2006}.
The chondritic porous subset of interplanetary dust particles (IDPs), which is most likely of cometary origin, shows that mineral grains are embedded in organic-rich carbonaceous material \citep[e.g.,][]{flynn-et-al2013}. 
The organic coating of mineral grains is similarly found in IDPs collected during the passage of Earth through the dust trail of Comet 26P/Grigg-Skjellerup \citep{busemann-et-al2009}.
Therefore, the 69~$\mu$m feature depends not only on the grain temperature and the iron fraction in olivine, but also on the volume fraction of organic material and the carbonization degree of the organic material \citep{kimura2014}.\footnote{Note that infrared spectral features of pure olivine grains are also incompatible with infrared spectroscopic observations of so-called ``olivine'' features \citep[e.g.,][]{fabian-et-al2001}.}
{\sl Accordingly, we conclude that dust mineralogy in debris disks needs revision in order to correctly interpret the infrared spectra of debris disks, unless debris-disk dust is totally dissimilar to comet dust. }
 
Other papers covered the following topics: a review of the Multi-Sphere T-Matrix light scattering codes for large aggregates of spheres by Ludmilla Kolokolova; numerical modeling of light-scattering cross sections for aggregates of spheres by Ryo Tazaki; numerical simulation on the giant-impact induced formation of hot debris disks by Hiroshi Kobayashi; adaptive-optics imaging observations of debris disks in the near-infrared by Zahed Wahhaj.

\subsection{Protoplanetary disks and star-forming regions}

Icy mantles on grains in the form of H$_2$O, CO, CO$_2$, and organic molecules are the repository for a significant amount of oxygen and carbon in star-forming regions and dark molecular clouds. 
With the advent of the Infrared Space Observatory and the Spitzer Space Telescope, our understanding of the formation and evolution of interstellar ices have been dramatically improved. 
Erika Gibb reviewed our current knowledge about ice mantles in various environments ranging from dark molecular clouds to protoplanetary disks to comets \citep{gibb2014}.
Special attention was paid to how ices evolve from initial freeze-out in dark clouds through the star formation process and potential incorporation into comets.
A comparison of molecular abundances between low-mass protostars and comets reveals noticeable differences in their abundances, suggestive of ice processing in the solar nebula \citep{oeberg-et-al2011}.
It is highly unlikely that interstellar ice mantles alone were incorporated into comets, since the grain cores of interstellar origin are not present in cometary dust \citep[cf.][]{keller-messenger2011}.
{\sl Therefore, the differences in protostellar and cometary ice abundances may indicate that protostellar ices sublimate and then re-condense in protoplanetary disks as icy grains.}

Other papers covered the following topics: theoretical and observational findings of dust evolution in protoplanetary disks by Til Birnstiel; 3-dimensional smoothed-particle hydrodynamics simulation of gas and dust in protoplanetary disks by Jean-Fran\c{c}ois Gonzalez \citep{gonzalez-et-al2015}; a comprehensive analysis of laboratory, theoretical, and numerical data on the surface energy of amorphous silica by Hiroshi Kimura; an experimental study on the formation of Fe$_3$C ultrafine grains by Chihiro Kaito; radiative transfer simulation of infrared and millimeter radiation from fluffy aggregates in protoplanetary disks by Koji Murakawa; numerical modeling of the mass ejected during aggregate--aggregate collisions by Koji Wada; a theoretical study on the effect of electric-field heating of plasmas on the ionization balance of dusty disks by Shoji Mori;
an analysis of the $11.2~\mu$m absorption feature in the spectra of star-forming regions by Tho Do-Duy \citep{doduy-et-al2015}; an analytical study on the grain surface chemistry with fluctuating temperatures by Emeric Bron; spectroscopic mapping observations of a high-mass star-forming region by Takashi Shimonishi; infrared photometry of massive star-forming regions by Yasuki Hattori \citep{hattori-et-al2015}; a theoretical study of the effect of grain growth on star formation in low-metallicity collapsing clouds by Gen Chiaki.

\subsection{Interstellar medium}

The evolution of amorphous carbon materials due to processing by UV radiation and shocks in the ISM is the key to best understanding the variations in the observed properties of interstellar dust, as claimed by \citet{jones2014} in the last meeting ({\sc Cosmic Dust \MakeUppercase{\romannumeral 6}}).
Takashi Onaka reviewed the results of the latest infrared observations of interstellar dust with the recent infrared satellites: Spitzer, Herschel, and AKARI \citep{onaka2014,onaka-et-al2015}.
AKARI observations of PAH emission features have clearly shown the variations in the size distribution and properties of carbonaceous dust due to processing by UV radiation or shocks. 
The aliphatic/aromatic ratios of PAHs are found to systematically decrease with the ionization fraction in Galactic star-forming regions of more than 100 samples, indicating that carbonaceous dust is aromatized in the ISM due to UV radiation accompanied by photo-fragmentation. 
In external galaxies, PAHs are significantly enhanced in intergalactic regions shocked by galactic superwind and galaxy merger, suggesting that PAHs most likely originate from shock fragmentation of large submicrometer-sized carbonaceous grains. 
While the observed PAH characteristics could be reasonably well understood in the theoretical framework of amorphous carbon evolution in the ISM, the reader is referred to the paper by \citet{kaneda-et-al2014} for more AKARI results on PAHs and carbonaceous dust.
It is worth noting that there are a couple of space missions probing the physical and chemical properties of interstellar dust.
However, the direct detection of small aromatic-rich carbonaceous interstellar grains in the Solar System is of great difficulty, if not impossible.
In comparison to large silicate grains, these small carbonaceous grains are subject to external forces in the solar radiation and magnetic fields and may not reach the inner Solar System, which is currently accessible to space missions.
{\sl Consequently, it is most likely that laboratory analyses of interstellar dust samples collected by Stardust do not provide a comprehensive picture of interstellar dust composition.}

Other papers covered the following topics: infrared to submillimeter observations of interstellar dust in dense clouds by Mika Juvela \citep{juvela2015}; a thorough survey of interstellar polarization data in optical, near-infrared, and millimeter wavelengths by Antonio Mario Magalh\~{a}es; an experimental development of C$_{60}$ production by Abel Brieva; a fitting of AKARI data on the $3.3~\mu$m emission band of PAH molecules by Mark Hammonds \citep{hammonds-et-al2015}; a detailed modeling of grain surface chemistry in the ISM by Franck Le Petit; a bottom-up computational modeling of silicate dust grains by Stefan Bromley; a modeling of all-sky dust map based on Planck, IRAS, and WISE data by Gonzalo Aniano; numerical simulation on the trajectory of interstellar dust in the heliosphere by Veerle Sterken; a search for micrometer-sized interstellar grains in infrared extinction curves by Aigen Li; a theoretical study of the ultraviolet to optical interstellar extinction in terms of the contribution of carbon dust by Ajay Mishra; fluid dynamics simulation of interstellar gas and dust by Tom Hendrix; dust emission modeling of infrared cirrus at high Galactic latitudes by Ajay Mishra.

\subsection{Evolved stars and supernovae}

A supernova (SN) is an explosion of a massive supergiant star at its latest stage and is known to seed stardust into the surrounding ISM.
Takaya Nozawa presented observational detections and theoretical modeling of dust formation in SNe and supernova remnants (SNRs) \citep{nozawa2014}.
Although the total dust mass produced by a single supernova is one order of magnitude higher than that of AGB stars, the mass supply rate of dust from SNe into the ISM is comparable to that of AGB stars.
SNe produce submicrometer-sized amorphous silicate grains in their ejecta due to rapid cooling, while oxygen-rich AGB stars with high-mass loss rates of $\dot{M} > 3\times{10}^{-8}~{M}_\odot$ produce crystalline silicate grains \citep{cami-et-al1998}.
Analyses of interstellar dust returned to Earth by the Stardust mission revealed the presence of micrometer-sized grains with a forsteritic olivine core and an amorphous silicate mantle in the local ISM \citep{westphal-et-al2014}.
Another micrometer-sized particle of the Stardust samples is possibly silicon carbide and the other submicrometer-sized grains seem to be characterized by aggregates of magnesium-rich, iron-bearing silicates.
We notice that micrometer-sized grains and submicrometer-sized grains in the Stardust interstellar dust samples do not contradict our understandings of condensates in AGB stars and SNe, respectively.
In contrast, a simple ``classical'' picture of interstellar dust such as a model by \citet{draine-lee1984} appears different, although carbonaceous grains might be filtered off at the heliospheric boundary.
{\sl As a result, a proper mixture of grains formed in SNe as well as oxygen-rich and carbon-rich AGB stars are required to satisfactorily model the interstellar dust populations, by taking into account their processing in the ISM.}

Other papers covered the following topics: near-infrared observations of the supernova remnant IC443 by Takuma Kokusho \citep{kokusho-et-al2015}; a laboratory study on the infrared spectra of wustite samples by Chiyoe Koike; laboratory infrared spectral in situ measurements of nucleating silicate nanoparticles by Shinnosuke Ishizuka.

\subsection{Galaxies}

Active galactic nuclei (AGN) are the central regions of galaxies where a supermassive black hole accretes gas and dust, and releases substantial energy as radiation.
Dust is the cornerstone of the unification theory of AGN, in which all AGN are essentially the same object but look different, depending on sightlines \citep{antonucci1993}.
Namely, much of the observed diversity arises from different viewing angles toward the central engine and a dusty toroidal structure around it. 
When the dusty torus is viewed face-on, both the central engine and the broad-line regions can be seen directly causing objects to appear as type 1 AGN; 
When the dusty torus is viewed edge-on, the anisotropic obscuration created by the torus causes objects to appear as type 2 AGN. 
Rachel Mason reviewed our current understanding of the chemical makeup of the dust torus, the destruction and formation of dust in the torus, and its possible roles in both inflow to and outflow from the nucleus \citep{mason2015}.
The $10~\mu$m silicate feature is naively expected to appear in emission and absorption toward type 1 and 2 objects, respectively, but there are outliers, namely, type 1 AGN with silicate absorption and type 2 AGN with silicate emission.
The distributions of the $10~\mu$m feature strength including the outliers are well explained if the torus does not have a smooth density distributions, but is a clumpy medium consisting of optically thick dusty clouds \citep{nikutta-et-al2009}.
The $10~\mu$m emission feature from the hot surfaces of directly illuminated clouds on the inner side of the torus is characteristic of amorphous silicate grains, even though the hottest grains are located in the vicinity of their sublimation zone.
{\sl This may indicate that silicate grains under harsh environments of AGN undergo amorphization in a much shorter timescale than crystallization.}

Other papers covered the following topics: Herschel observations of cold dust in galaxies by Aur\'{e}lie R\'{e}my-Ruyer; AKARI observations of CO$_2$ and H$_2$O ices in nearby galaxies by Mitsuyoshi Yamagishi; Spitzer and Herschel observations of the Magellanic Clouds by Fr\'{e}d\'{e}ric Galliano; theoretical modeling of destruction and formation processes on carbonaceous grains in H\,{\sc ii} complexes by Maria Khramtsova; numerical simulation of dust evolution in star-forming galaxies by Kenji Bekki; AKARI observations of PAHs in anomalous microwave excess regions by Aaron C. Bell; a theoretical study of the dust-to-metal ratio in galaxies by Akio K. Inoue.

\section{Perspectives for the development of cosmic dust research}
\label{perspectives}

The rationale behind the meeting is a development of cosmic dust research by an interdisciplinary approach to answer the questions of where dust comes from and where dust goes \citep[see][]{kimura-et-al2014}.
We are pleased that several presentations in {\sc Cosmic Dust \MakeUppercase{\romannumeral 7}} extended the studies across different environments of dust and ices in terms of their evolution and processing \citep[e.g.,][]{olofsson-et-al2014,gibb2014}.
These are promising works in the development of cosmic dust research, as an interdisciplinary research is essential in achieving great scientific advances in the cosmic dust field.

By looking back on {\sc Cosmic Dust \MakeUppercase{\romannumeral 7}}, we could link several presentations together to gain a new insight into the formation and evolution of dust and ices.
The surfaces of dust particles provide a vital role in the formation of major molecules observed in a variety of environments such as cometary comae, protoplanetary disks, the interstellar medium, and galaxies.
The abundances of CH$_3$OH, CH$_4$, and CO with respect to H$_2$O in low-mass protostars are clearly higher than those in comets, which is so far interpreted by processing in protoplanetary disks \citep{gibb2014}.
The average abundance of solid CO$_2$ relative to solid H$_2$O in low-mass protostellar envelopes is close to 0.3, which is higher than the value around $0.2$ found in dense molecular clouds and massive star-forming regions \citep{pontoppidan-et-al2008}.
It turned out in the meeting that the ratios of CO$_2$ to H$_2$O ice abundances in comets and nearby galaxies lie in the range of 0.05--0.30 \citep{ahearn2014,yamagishi-et-al2014}.
Once CO$_2$ molecules are created either on the surface of refractory grains or in the gas phase, the molecules are stable against chemical reactions that could reduce the abundance of the molecules.
\citet{yamagishi-et-al2014} have shown that the variation of the CO$_2$/H$_2$O ratio in nearby galaxies seems to result from its dependence on the irradiation of ultraviolet photons induced by cosmic rays.
If the CO$_2$/H$_2$O ratio is a good tracer for the hardness of the UV radiation field, the CO$_2$/H$_2$O ratio in comets might provide a clue about the location where comets were formed in the solar nebula.
However, the discovery of pure icy grains in comets may violate the common belief that H$_2$O and CO$_2$ molecules cannot be efficiently formed in the gas phase and thus the surface of dust acts as a catalyst for the formation of these molecules.
It is, therefore, essential to rethink the formation of icy grains in protoplanetary disks from different angles, which will hopefully be addressed in the next meeting.

We cordially invite the reader to take part in the 8th meeting on Cosmic Dust ({\sc Cosmic Dust \MakeUppercase{\romannumeral 8}})\footnote{{Contact:~}{dust-inquries@cps-jp.org}} and to join us for the development of cosmic dust research.
{\sc Cosmic Dust \MakeUppercase{\romannumeral 8}} will be held at the Tokyo Skytree Town Campus of Chiba Institute of Technology, Tokyo, Japan on August 17--21, 2015.
The deadline for admissions application is May 13, 2015, but early birds who complete both admissions application and abstract submission by April 30, 2015 will receive a discount on the registration fee.
More information on {\sc Cosmic Dust \MakeUppercase{\romannumeral 8}} is available at the Cosmic Dust website ({https://www.cps-jp.org/\~{}dust}).

\section*{Acknowledgements}
\label{acknowledgements}

We express our sincere gratitude to Hiroki Chihara (LOC Chair), Koji Wada, Hiroshi Kobayashi, Hiroki Senshu, Takashi Shimonishi, Ryo Tazaki, and Takayuki Hirai for their cheerful dedication to the organization of {\sc Cosmic Dust \MakeUppercase{\romannumeral 7}} as members of the LOC (Local Organizing Committee).
We thank all the authors and the reviewers as well as the editorial board of PSS and Elsevier for putting their efforts into this special issue. 
We are indebted to Osaka Sangyo University, JSPS (Japan Society for the Promotion of Science), and Society for Promotion of Space Science ({http://www.spss.or.jp}) for their various support for organizing the meeting.

\bibliographystyle{model2-names}
\bibliography{<your-bib-database>}



\begin{flushright}
Hiroshi Kimura\\
{\sl Graduate School of Science, Kobe University, c/o CPS (Center for Planetary Science), Chuo-ku Minatojima Minamimachi 7-1-48, Kobe 650-0047, Japan}

\vspace{0.2in}
Ludmilla Kolokolova\\
{\sl Planetary Data System Group, Department of Astronomy, University of Maryland, College Park, MD 20742, USA}

\vspace{0.2in}
Aigen Li\\
{\sl 314 Physics Building, Department of Physics and Astronomy, University of Missouri, Columbia, MO 65211, USA}

\vspace{0.2in}
Jean-Charles Augereau\\
{\sl Universit\'e Grenoble Alpes, IPAG, F-38000 Grenoble, France}\\
{\sl CNRS, IPAG, F-38000 Grenoble, France}

\vspace{0.2in}
Hidehiro Kaneda\\
{\sl Graduate School of Science, Nagoya University, Furo-cho, Chikusa-ku, Nagoya 464-8602, Japan}

\vspace{0.2in}
Cornelia J\"{a}ger\\
{\sl Max Planck Institute for Astronomy, Heidelberg, Laboratory Astrophysics and Clusterphysics Group at the Institute of Solid State Physics, Friedrich Schiller University Jena, Helmholtzweg 3, 07743 Jena, Germany}
\end{flushright}

\end{document}